\documentclass[structabstract]{aa}  
\usepackage{graphicx}
\usepackage{txfonts}
\usepackage{natbib}
\usepackage{deluxetable}
\begin{document}
   \title{Searching for the 511 keV annihilation line from galactic compact objects with the IBIS gamma ray telescope}

   \author{G. De Cesare
          \inst{1}
          }

   \institute{INAF-Istituto di Astrofisica Spaziale e Fisica Cosmica di Roma, via Fosso del Cavaliere 100, I-00133  Roma, Italy\\
              \email{giovanni.decesare@iasf-roma.inaf.it}       
             }

   \date{Received ...; accepted ...}

 
  \abstract
   {The first detection of a gamma ray line with an energy of about 500~keV from the center of our Galaxy dates back to the early seventies. Thanks to the astrophysical application of high spectral resolution detectors, it was soon clear that this radiation was due to the 511~keV photons generated by electron-positron annihilation. Even though the physical process are known, the astrophysical origin of this radiation is still a mystery.}
   {The spectrometer SPI aboard the INTEGRAL gamma-ray satellite has been used to produce the first all-sky map in light of the 511~keV annihilation,  but no direct evidence of any 511~keV galactic compact objects has been found. Owing to its moderate angular resolution, these SPI data are still compatible with a distribution of point sources clustered in the bulge of our Galaxy. Thanks to the fine angular resolution and the large field of view the IBIS imager on the INTEGRAL satellite gives us the unique opportunity to search for a possible 511~keV line from point sources associated to known objects, such as X-ray binaries, or supernovae, or even new ones.}
   {We present the first deep IBIS 511 keV all sky map, obtained by applying standard analysis to about 5 years of data. Possible 511~keV signals are also searched over hour-day-month timescales. The IBIS sensitivity at 511~keV depends on the detector quantum efficiency at this energy and on the background. Both these quantities were estimated in this work.}
   {We find no evidence of Galactic 511 keV point sources.  With an exposure of 10~Ms, in the center of the Galaxy, we estimate a $1.6 \times 10^{-4}\,ph\,cm^{-2}\,s^{-1}$ flux 2 sigma upper limit. Similar limit is given in a wide area in the Galactic center region with similar exposures. The IBIS 511~keV flux upper limits for microquasars and supernova remnants detected in the hard X domain ($E >  20\, keV$) are also reported.}
   {Our results are consistent with a diffuse $e^{+}e^{-}$ annihilation scenario. If positrons are generated in compact objects, we expect that a significant fraction of them propagate in the interstellar medium before they are annihilated away from their birth places.}


   \maketitle
%

\section{Introduction}
Positrons can be observed through the detection of gamma rays produced by the electron-positron annihilation. This process can proceed in two ways, by direct annihilation or by the formation of positronium, i.e. bound states consisting of an electron and a positron. Twenty-five percent of the positronium atoms have antiparallel spins (para-positronium, p-Ps) and 75 \% have parallel spins (ortho-positronium, o-Ps). It is known that p-Ps has a lifetime of $1.25 \times 10^{-10}\, s$ and decays into two gamma rays each with 511 keV energy, while o-Ps has a lifetime of $1.5 \times 10^{-7}\, s$ and three gamma rays are emitted, the maximum energy being 511 keV. By evaluating the ratio between the line and the continuum component in the spectrum, one can estimate the fraction of positrons that decay by formation of positronium.

   The discovery of positrons in our Galaxy dates back to the seventies, when in a balloon experiment \cite{Johnson1972} detected gamma ray line emission at the energy of $476\pm 26$ keV from the Galactic center. However, due to the low energy resolution of the detectors, the physical origin of this emission was unclear. A few years later, thanks to the advent of high-resolution spectrometers, this radiation was identified as a 511~keV narrow line ($FWHM < 3\, keV$) produced by electron-positron annihilation \citep{Leventhal1978}. Observations of the galactic annihilation radiation continued progressively until today with the new observations made with the INTErnational Gamma-Ray Astrophysics Laboratory (INTEGRAL). The INTEGRAL satellite,  launched from Baikonur in Kazakhstan on October 17, 2002, is an ESA mission in cooperation with Russia and the United States.
   
   The 511~keV flux ($\sim 10^{-3}\, ph\, cm^{-2}\, s^{-1}$) measured by the spectrometer SPI  aboard the INTEGRAL satellite shows that $\sim\, 10^{43}$ positrons per second  annihilate in the bulge of our Galaxy \citep{Weidenspointner2008a}. The origin of this large number of positrons is unclear. Indeed, although the physical processes responsible for producing positrons are quite well understood, it is hard to find a picture that explains both this high rate of annihilation and the positron distribution in the Galaxy.  The important observational constraints obtained so far can be divided into three main categories: (1) the spectral 511~keV line properties that give information on the environment where the positrons annihilate; (2) the spatial distribution of the emission  \citep[see e.g.][]{Weidenspointner2008}, which must be correlated with the possible positron sources candidates; and (3) the continuum galactic soft $\gamma$-ray emission, which constrains the positron injection energy. Concerning points (1) and (2), the SPI data show that the 511~keV line width is on the order of keV, and the emission, contrary to what happens at other wavelengths, is mainly coming from the Galactic bulge, in a region spatially extended on a scale of about 8 degrees. The measured ortho-positronium continuum flux yields a fraction of positronium of $(96.7 \pm 2.2)\, \%$ \citep{Jean2006}. Assuming a Galactic center distance of 8.5~kpc, the bulge size corresponds to approximately 1.2~kpc. Concerning point (3), the lack of detection of a $\gamma$-ray continuum by COMPTEL/EGRET due to in-flight annihilation puts a severe constraint on the positrons injection energy. A calculation of the $\gamma$-ray spectrum produced by the positron annihilation while they are still relativistic has shown that the emerging photon flux would exceed the COMPTEL/EGRET flux limit unless the positrons are injected into the interstellar medium with an energy below $\sim$~3~MeV \citep{Beacom2006}. An additional observational constraint is the flux upper limit on possible point sources. 
   Possibly due to positron annihilation, transient broad-line emission has been reported in a few cases. High-energy emissions characterized by a broad line in the  300-600~keV range from 1E~1740.7-2942 \citep{Sunyaev1991, Bouchet1991} and Nova Muscae \citep{Gilfanov1991} have been detected by the imaging $\gamma$-ray telescope SIGMA aboard the GRANAT spacecraft.  These events were interpreted as arising from positron annihilation near a galactic black hole  \citep{Ramaty1981}; however the 1E~1740.7-2942 1-day emission in September 1992 was not confirmed by OSSE  \citep{Jung1995}. We note that the different physical characteristics of the environment inferred from the line width mean there is no simple connection between these observations and the diffuse 511~keV galactic emission.  

   Several scenarios have been proposed to explain observations. In general, because of the poor angular resolution of SPI, we cannot exclude the possibility that the Galactic 511~keV gamma-ray line emission originates in a limited number of point sources \citep{Knodlseder2005}.  However, the spectral properties of the 511~keV line measured by SPI \citep{Jean2004} suggest that a large fraction of positrons generated in compact objects should propagate in the interstellar medium before being annihilated. The SPI data can be explained in terms of the annihilation of positrons in a static warm ($\sim 10^{4}\, K$) or cold ($\sim 10^{2}\, K$) slightly  (a few \%) ionized medium or, as has been recently proposed by \cite{Churazov2010}, with a model of positron annihilation in a radioactively cooling interstellar medium. Among the compact objects, we know that the supernovae must emit positrons by the $\beta^+$ decay of the nuclear synthesis products. One problem in this case is that the bulk of positrons are observed from the Galactic bulge, in contrast to what we would expect from the supernovae distribution in the Galactic plane. A possible solution has been proposed by \cite{Higdon2009}, who suppose that a large amount of positrons generated in the Disk by the supernovae explosions are transported to the bulge following the galactic magnetic field lines. The low mass X-ray binaries (LMXBs) have been proposed by  \cite{Weidenspointner2008} as a class of positron sources  that could explain the 511~keV emission line measured by INTEGRAL. Basically, this idea is supported by two arguments. First, the LMXBs are old systems concentrated in the Galactic bulge, where we detect the main amount of annihilation radiation. The second reason is that the SPI results show an asymmetric 511 keV emission morphology with a spatial shape that is consistent with the LMXBs distribution above 20 keV. In X-ray binaries, positrons in $e^{+}e^{-}$ pairs can be created in the vicinity of the compact object, either in the hot inner accretion disk, in the X-ray corona surrounding the disk, or at the base of the jet, which may channel a fraction of $e^{+}e^{-}$ pairs out of the system \citep{Prantzos2010}.  Finally, the annihilation can be also explained by new astrophysical ideas. The possibility that the annihilation radiation is due to the $e^{+}e^{-}$ pairs produced when X-rays generated from the Galactic center black hole interact with $\sim\, 10\, MeV$ temperature blackbody low-mass ($10^{17}\, g$) black holes within $10^{14}$ -- $10^{15}\, cm$ of the center is discussed by \cite{Titarchuk2006}. Annihilation of light dark matter (DM) particles into $e^{+}e^{-}$ is also a possible explanation \citep{Boehm2004}.
   
In this paper we use IBIS (the $\gamma$-ray Imager on Board the INTEGRAL Satellite) to estimate the 511~keV flux upper limits on possible Galactic point sources. In \S \ref{data-analysis} we describe the data analysis, and in \S \ref{results} and \S \ref{discussion} we present and discuss our results.
   

\begin{figure*}[t]
\centering
\includegraphics[width=14cm,height=6cm]{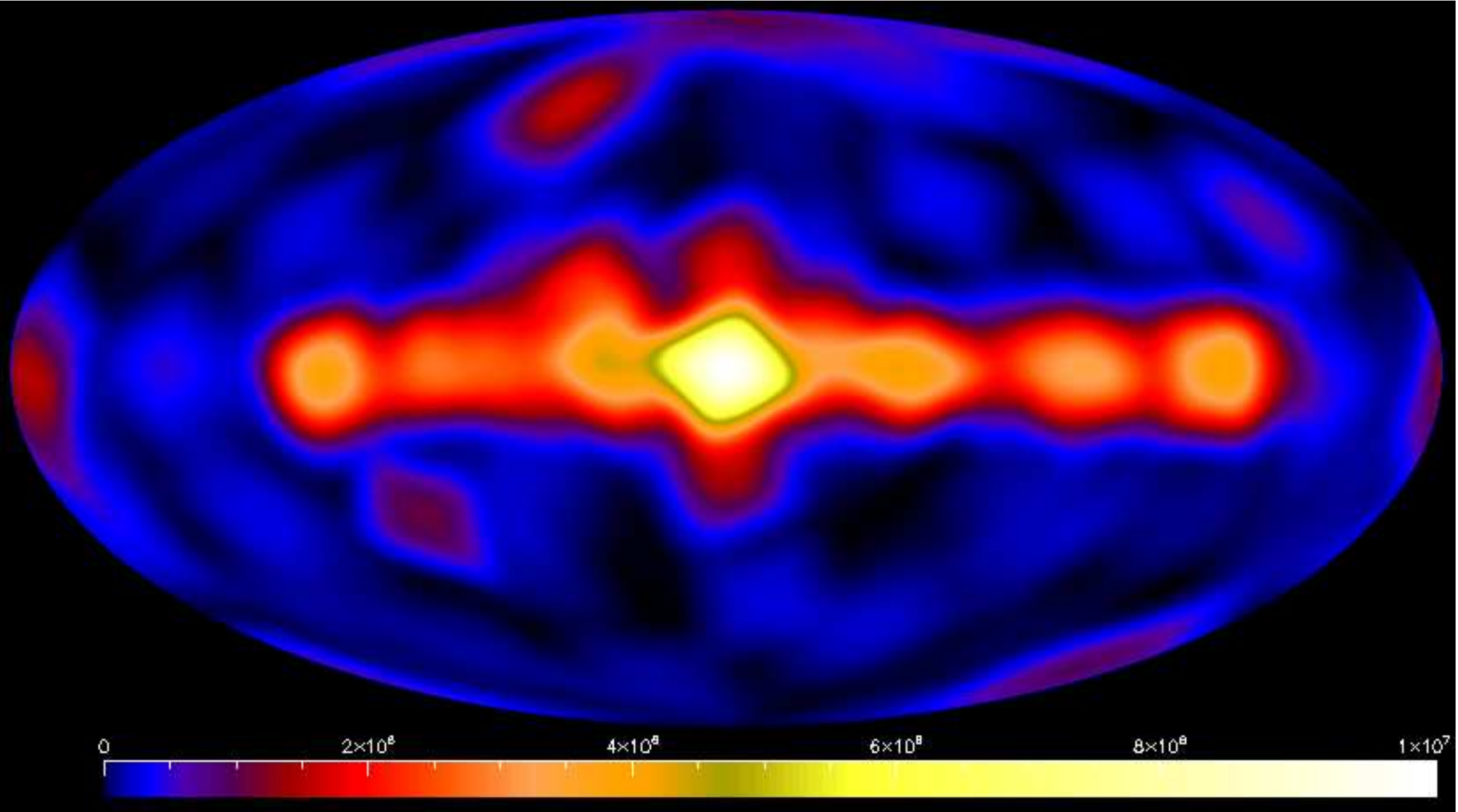}
\caption{The exposure map for our data set. The deeper observation correspond to the galactic plane and the Galactic Center, where we reach with our data a 10 Ms exposure}
\label{fig:exp}
\end{figure*}
   
\section{Data analysis}
\label{data-analysis}
\subsection{Data sample}
      The last session of our IBIS 511~keV data reduction started in April 2008. Our data set consists of all the IBIS public data available at that time, i.e. about 5 years of observations, from October 17, 2002, when INTEGRAL was launched, until April 2007. In addition, we used all the Core program data until April 2008. The data sample, which corresponds to the 4th IBIS hard X sky survey \citep{Bird2010} has been filtered to exclude the observations affected by strong background and systematic noise (typically due to solar flares). 
      
       All the data that we reduced correspond to 39413 IBIS pointings,  also called science windows ({\em ScWs}), each lasting about $2000\, s$. The total observing time in our data sample is $\sim\, 80\, Ms$.  The sky exposure map we obtain with this data set is shown in Figure \ref{fig:exp}. We note that the maximum exposure (about 10 Ms) corresponds to the Galactic center region, the region where the bulk of the positron emission has been detected by the SPI spectrometer aboard the INTEGRAL satellite.

\subsection{Standard analysis}
     The IBIS telescope is based on the use of the coded aperture mask technique \citep{Caroli1987}, that is a key  method for providing images at energies above tens of keV, where photon focusing becomes impossible using standard grazing techniques. This established technique is remarkably successful for weak extragalactic sources, as well as for crowded galactic regions like the Galactic center. In a coded mask telescope, the incoming source photons are, before detection, encoded by a mask, so the image of the sky has to be reconstructed by decoding the observation afterward. The coded masks optimize the background subtraction capability because of the opportunity to observe the source and the off-source sky at the same time. In fact, for any particular source direction, the detector pixels are divided into two sets: those capable of viewing the source and those for which the flux is blocked by opaque mask elements.  The coded mask telescope are also characterized by a very large field of view (FOV). The FOV is divided in two parts, the fully coded (FC) FOV for which all source radiation directed towards the detector plane is modulated by the mask and the partially coded (PC) FOV for which only a fraction of it is modulated by the mask. 

     The IBIS data and software are delivered by the INTEGRAL Scientific Data Center (ISDC) in Geneva \citep{Courvoisier2003}. The Off-line Scientific Analysis (OSA) software is regularly released by ISDC with bug fixes and some improvement, we used the release 7.0 of this software  in our data analysis.  The scientific data analysis methods,  hereafter called {\em Standard Analysis}, are reported by \cite{Goldwurm2003a}. In this work the IBIS data have been reduced using the IASF-INAF computing facility in Rome \citep{Federici2009}. 

\subsection{The IBIS Monte Carlo}
\label{sec:mc}
The first motivation behind the effort by the community of IBIS in developing the Monte Carlo is to calculate the detector {\em response matrix}. The  {\em response matrix} is a table that determines the probability $p(E,k)$ that a photon of energy $E$, incident on the detector plane, produces a signal in the detector channel $k$. The IBIS modeling activities  \citep{Laurent2003} were performed step by step by firstly constructing a {\em calibrated model} of each IBIS sub-system: ISGRI, PICsIT, Mask, active shield, etc. A  {\em calibrated model}  is defined as a GEANT Monte Carlo model, further checked by intensive corresponding calibration. The calibrated models have been integrated in a second step in the whole IBIS mass model. This mass model was then checked using  the data obtained during the IBIS telescope calibration \citep{Bird2003} in the payload ground calibration at ESA/ESTEC, and refined using in-flight calibration data \citep{Terrier2003} acquired during the whole INTEGRAL mission.

In this work, we apply the Monte Carlo to the evaluation of the detector effective area at 511~keV, in order to convert the detector counts in photons per unit of area.

\subsection{Evaluation of the 511 keV line flux upper limit}
      The $\gamma$-ray {\em sensitivity} is defined as the minimum flux detectable at a level that can be distinguished from the noise;  the lower the {\em sensitivity}, the higher the signal-to-noise ratio for a given source. The {\em sensitivity} depends on the significance level, so we need to define at what level it becomes a reliable signal coming from a real source. In this work we assume that we get a positive signal if it is measured at a significance level greater than $2\,\sigma$.
      
      Irrespective of the imaging system we use, the detector count rate $C$ is the sum of the source $S$ and the instrumental background $B$. Then
\begin{equation}
S = C - B
\label{eq:signal}
\end{equation}
For any given source, in a coded mask system, the source and background counts can be measured using the detector areas that correspond to open and closed mask elements, respectively. In terms of count rate, the astrophysical measures in the hard X and $\gamma$-ray domains are dominated by the background ( $B \gg S$). In the {\em Standard Analysis}, assuming a Poissoniam noise, the variance in each reconstructed sky pixel of the FCFOV is given by the total count recordered by the detector \citep{Goldwurm2003a}. Thus, if a source produces the count rate $S$ on the detector, for a given exposure $T$, the signal to noise ratio is simply
\begin{equation}
\frac{s}{n} = \frac{Source\; Count}{\sqrt{Total\; Count}} =\frac{S \times T}{\sqrt{(S + B) \times T}} \simeq S \times \sqrt{\frac{T}{B}}
\label{eq:sens}
\end{equation}

The source signal $S$ depends on the detector area and on the {\em imaging efficiency}; i.e., for a given source, the fraction of detected photons that are effectively used to form the signal $S$. In general, due to finite detector space resolution and the mask transparency, this number is lower than one. More precisely, the {\em imaging efficiency} depends on the ratio between the mask pixel and the detection units size (the more pixels, the higher the imaging efficiency) and on the real transparency of the open and closed mask elements. The mask thickness of $16\, mm$ guarantees a very low transparency, $t_{0} = 3\, \%$ of the close elements at 511~keV, and a high transparency, $t_1= 90\, \%$ of the open elements. The measured values of the open-mask element transparency versus the incident angle are reported for many source energies by  \cite{Sanchez2005}: at energies higher than $\sim$ 30~keV, in particular at 511~keV, the transparency has a low dependence on the incident angle.The imaging efficiency $eff_{img}$ can be evaluated by the expression \citep{Skinner2008}
\begin{equation}
eff_{img} =
\left\{
\begin{array}{rl}
 \left(1 - \frac{d}{3 m}\right)(t_1-t_0) \sqrt{4f(1-f)} & \mbox{if } d < m \\
 \frac{m}{d} \left(1 - \frac{m}{3 d}\right)(t_1-t_0) \sqrt{4f(1-f)} & \mbox{if } d \geq m
\end{array}
\right.
\label{eq:imaging}
\end{equation} 
where the mask open fraction $f$ is 0.5 for IBIS, $d$ and $m$ are the size of the detector and mask units.  For IBIS/ISGRI, because $d/m = 4.6/11.2 = 0.4107$, taking the mask's closed and open element transparencies $t_{0}$ and $t_{1}$ into account, the imaging efficiency is
\begin{equation}
eff_{img} (511\, keV) =  0.75
\label{eq:isgri-img-eff}
\end{equation}




\section{Results}
\label{results}
\begin{figure*}
\centering
\includegraphics[width=13.4cm,height=8cm]{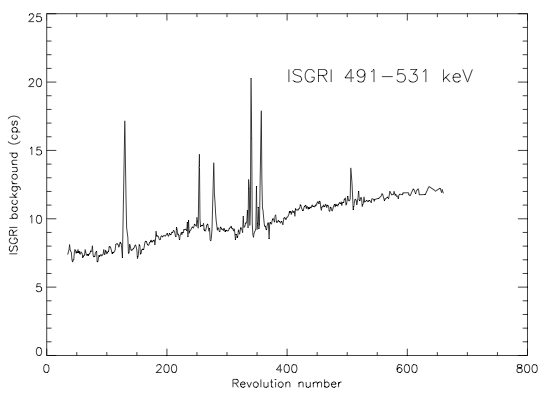}
\caption{The 511 keV line background as measured by ISGRI. The energy band width is 1 FWHM of the detector energy resolution at 511 keV. Each INTEGRAL revolution in the horizontal axis lasts 3 days.}
\label{fig:bkg}
\end{figure*}

\begin{figure*}
\centering
$\begin{array}{ccc}
\includegraphics[width=7.5cm,height=7.5cm]{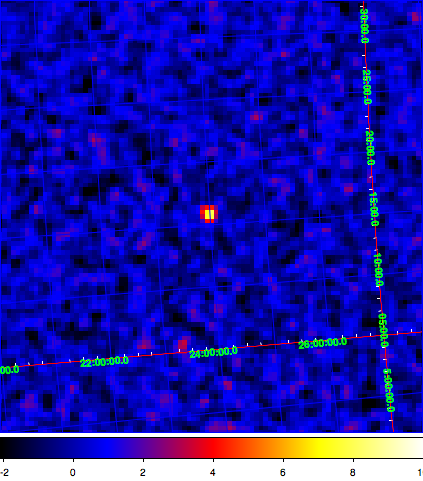} &
\includegraphics[width=0.2cm,height=7.6cm]{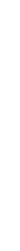} &
\includegraphics[width=7.5cm,height=7.5cm]{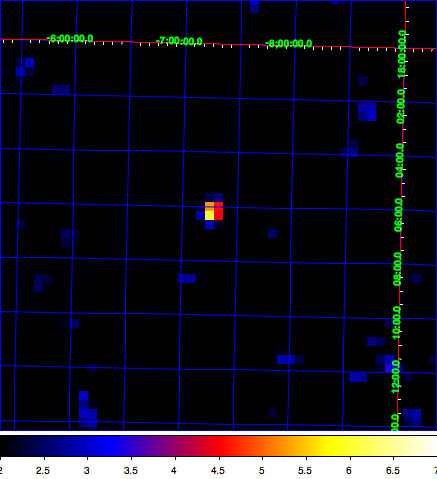} 
\end{array}$
\caption{A simulation of a 511~keV point source as seen by IBIS/ISGRI during a pointing in the early (rev. 50), left panel, and late (rev. 600), right panel, orbits of the INTEGRAL satellite. The signal level of the detected 511~keV source goes to $10\, \sigma$ in the first image to $7\, \sigma$ in the second one. These images have been obtained adding to the real data the data generated by the IBIS Monte Carlo and then processing the result with the standard analysis software until the image level.}
\label{fig:mc-and-data}
\end{figure*} 
\begin{figure*}[t]
\centering
$\begin{array}{cc}
\includegraphics[width=7.cm,height=7.cm]{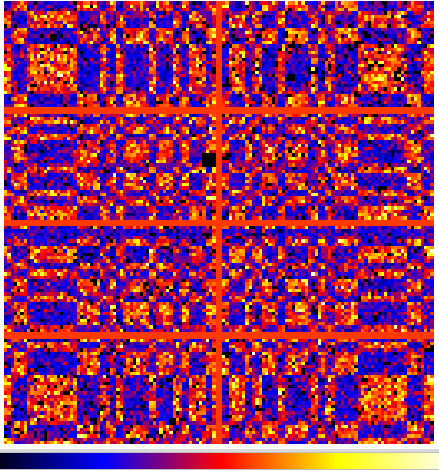} &
\includegraphics[width=7.cm,height=7.cm]{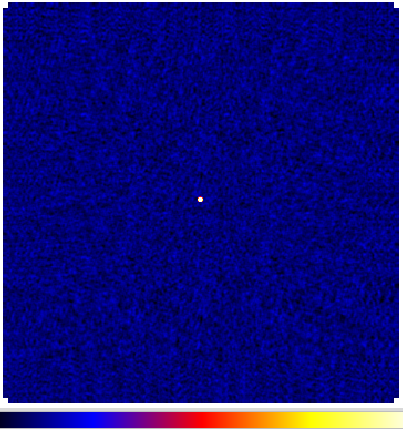} \\
\includegraphics[width=7.cm,height=7.cm]{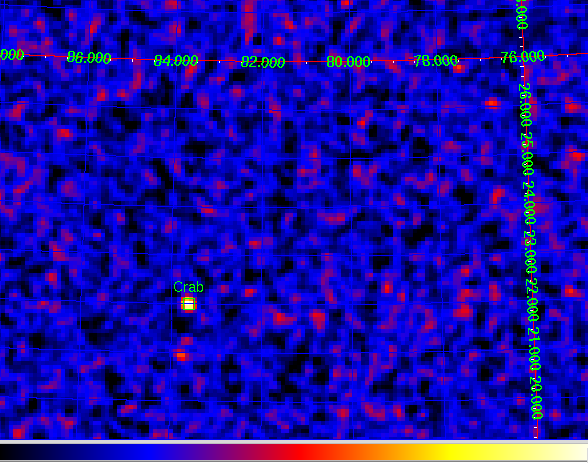} &
\includegraphics[width=7.cm,height=7.cm]{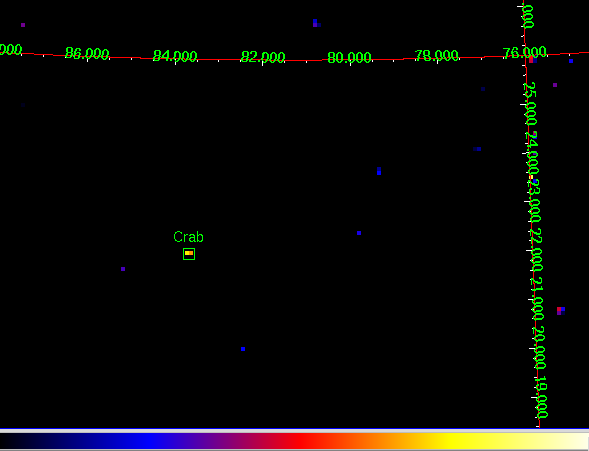}
\end{array}$
\caption{The Crab nebula as detected by IBIS/ISGRI in the soft $\gamma$-ray domain with a 3~Ms of exposure. Left and right top panel: the ISGRI 40-100~keV {\em shadowgram} accumulated during one ScW (no. $010200210010$) and the decoded sky image. Left bottom panel: sky mosaic in $431$-$471$ keV (colorbar: $-2$ - $8\, \sigma$). Right bottom panel:  sky mosaic in $491$-$531$ keV (colorbar: $3$ - $4\, \sigma$).}
\label{fig:crab-gamma}
\end{figure*} 
\begin{table}
\caption{The Crab nebula as observed by IBIS with an exposure of $3\, Ms$.}
\begin{center}
\begin{tabular}{c|c|c}
\hline\hline
Energy (keV) & signal ($\sigma$) & counts/s \\
\hline
$20-40$ keV      &  8372       & $157.527 \pm 0.019$      \\
$40-100$ keV    &  5360       &  $93.984  \pm  0.017$      \\
$100-300$ keV  &  1250       &  $17.608 \pm 0.014$         \\
$431-471$ keV  &     9.5       &  $(3.0 \pm 0.3)\, 10^{-2}$     \\
$491-531$ keV  &     3.8       &  $(1.2 \pm 0.3)\, 10^{-2}$  \\
$551-591$ keV  &     not detected       & $-$     \\
\hline
\end{tabular}
\end{center}
\label{tab:crab}
\end{table}
\begin{table}[t]
\begin{center}
\begin{tabular}{|l|c|}
\hline
parameter & value \\
\hline
\hline
photon index for $E < E_{break}$                         & $2.08 \pm 0.01$      \\
$E_{break}$  (keV)                                               &  $100$    (frozen)  \\
photon index for $E > E_{break}$                         & $2.23    \pm 0.03$   \\
photons $cm^{-2}\, s^{-1}\, keV^{-1}$ at 1 keV   &   $9.3     \pm 0.2$   \\
\hline
\end{tabular}
\end{center}
\caption{The broken power law model for the Crab spectrum in X- and soft $\gamma$-rays reported by \cite{Jourdain2008}}
\label{tab:broken}
\end{table}%

\begin{table*}
\caption{\label{t7}511 keV $2\,\sigma$ Sgr A* upper flux limit in the Galactic Center visibility periods.}
\centering
\begin{tabular}{l c c c c c}
\hline\hline
Period & Start Time & End Time & No. ScWs\tablefootmark{a} & Sgr A* Exposure\tablefootmark{b} & Flux limit \\
            & (UT)            & (UT)          &                                                 & Ms                                                        &   $10^{-4}\,ph\,cm^{-2}\,s^{-1}$ \\ 
\hline
 2003 Spring & 2003-02-28 & 2003-04-23 & 1731 & 0.650 & 5.7 \\
 2003 Autumn & 2003-08-10 & 2003-10-14 & 1717 & 1.872 & 3.2 \\
 2004 Spring & 2004-02-16 & 2004-04-20 & 1862 & 1.046 & 4.5 \\
 2004 Autumn & 2004-08-17 & 2004-10-27 & 2191 & 1.292 & 4.3 \\
 2005 Spring & 2005-02-16 & 2005-04-28 & 2144 & 1.165 & 4.7 \\
 2005 Autumn & 2005-08-16 & 2005-10-26 & 1667 & 0.790 & 5.9 \\
 2006 Spring & 2006-02-09 & 2006-04-25 & 1869 & 1.501 & 4.1 \\
 2006 Autumn & 2006-08-16 & 2006-11-02 & 1770 & 1.080 & 5.0 \\
 2007 Spring & 2007-02-01 & 2007-04-22 &  985 & 0.347 & 9.2 \\
 \hline
\end{tabular}
\tablefoot{ \\
\tablefoottext{a}{Each ScW lasts about half an hour.}\\
\tablefoottext{b}{Obtained from the exposure map. The exposure map, and in particular the value in the Galaxy depends both on the number of ScWs and the dithering patterns}
}
\label{tab:gcvis}
\end{table*}

\begin{figure}
\centering
\includegraphics[width=9cm,height=7cm]{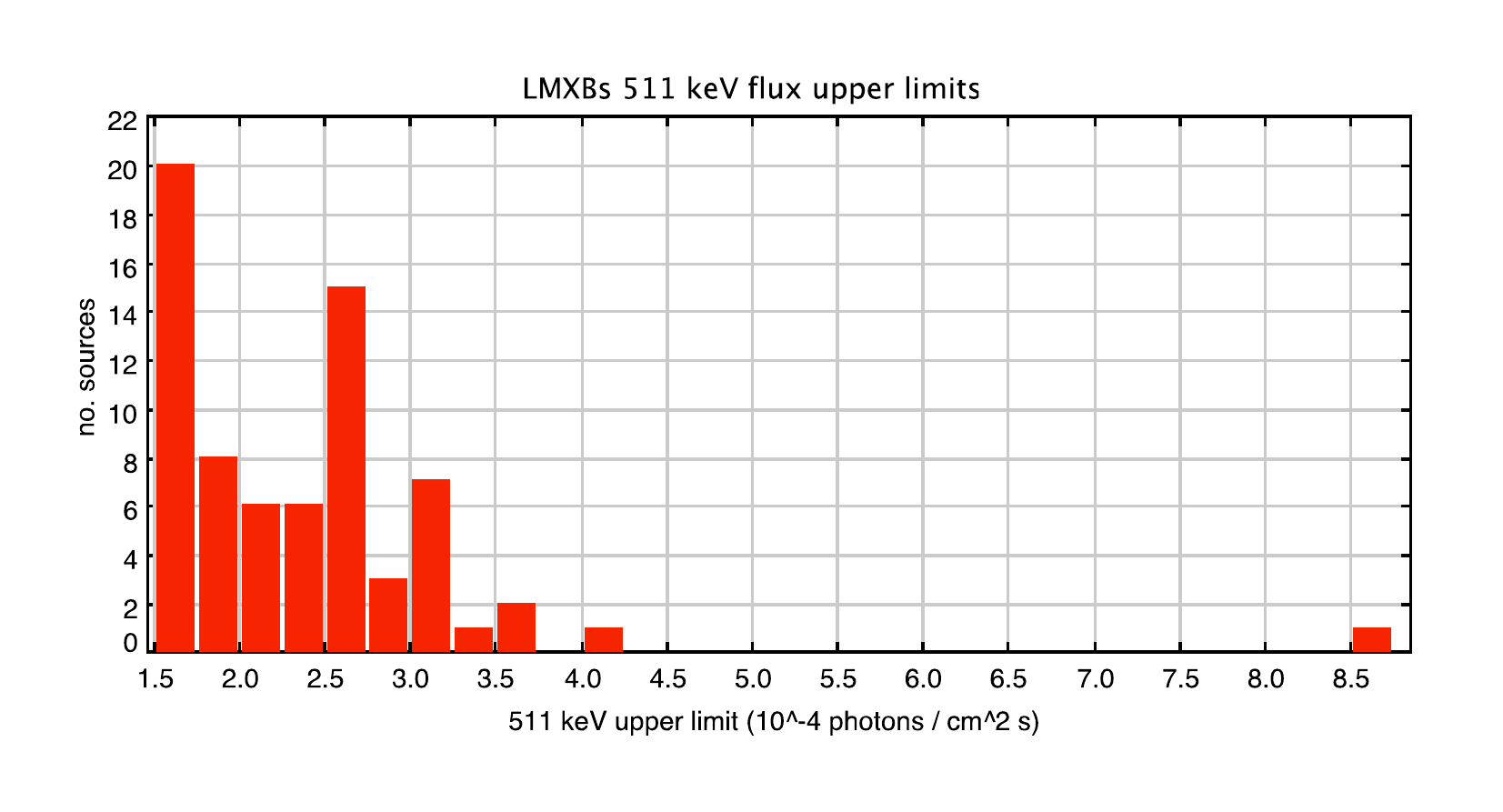}
\caption{The distribution of 511 keV flux upper limits for the IBIS LMXBs population. The first peak correspond to a $2\sigma$ upper limit of $1.6 \times 10^{-4}\,ph\,cm^{-2}\,s^{-1}$, estimated for the IBIS sources located in the Galactic Bulge.}
\label{fig:lmxb-hist}
\end{figure}

\begin{table*}
\caption{\label{t7}511 keV $2\,\sigma$ flux upper limits for the hard X (E $>$ 20~keV) microquasars reported in the 3rd IBIS catalogue \citep{Bird2007}.}
\centering
\begin{tabular}{l r r c r r c c}
\hline\hline
Source name & R.A. & Dec & Error\tablefootmark{a} &  $F_{20-40}$\tablefootmark{b} & $F_{40-100}$\tablefootmark{b} & $2\,\sigma$ flux limit &  Type\tablefootmark{c} \\
                      & (deg)  & (deg)  &  (arcminutes)   &  (mCrab)         &  mCrab            & ($10^{-4}\,ph\,cm^{-2}\,s^{-1}$) \\
\hline
XTE J1550-564  & 237.745 & -56.479  & 0.2 & $34.0 \pm 0.1$  & $55.3 \pm 0.2$ & 2.7 & LMXB \\       

Sco X-1        & 244.980 & -15.643  & 0.2 & $685.7 \pm 0.3$ & $24.7 \pm 0.3$ & 3.7 & LMXB  \\        

GRO J1655-40   & 253.504 & -39.846  & 0.6 & $2.3 \pm 0.1$   & $2.7 \pm 0.2$  & 2.5 & LMXB  \\        

GX 339-4       & 255.706 & -48.792  & 0.3 & $40.7 \pm 0.1$  & $46.7 \pm 0.2$ & 2.6 & LMXB  \\       

1E 1740.7-2942 & 265.978 & -29.750 & 0.2 & $29.8 \pm 0.1$  & $36.6 \pm 0.1$ & 1.6 & LMXB  \\       

GRS 1758-258   & 270.303 & -25.746  & 0.3 & $58.8 \pm 0.1$  & $75.3 \pm 0.1$ & 1.6 & LMXB  \\         

SS 433         & 287.956 & 4.983    & 0.5 & $10.4 \pm 0.1$  & $5.2 \pm 0.2$  & 2.4 & HMXB  \\      

GRS 1915+105   & 288.799 & 10.944 & 0.2  & $296.8 \pm 0.1$ & $123.4 \pm 0.2$  & 2.5 & LMXB  \\      

Cyg X-1        & 299.590 & 35.199   & 0.2 & $763.7 \pm 0.2$ & $876.7 \pm 0.3$  & 3.0 & HMXB  \\       

Cyg X-3        & 308.108 & 40.956   & 0.2 & $196.5 \pm 0.2$ & $78.3 \pm 0.3$   & 2.6 & HMXB \\ 
\hline
\end{tabular}
\tablefoot{\\
\tablefoottext{a}{Position error expressed as radius of 90 \% confidence circle in arcminutes.} \\
\tablefoottext{b}{Fluxes averaged on the total exposure: \\
      20-40 keV: $1\,mCrab = 7.57 \times 10^{-12}\,erg\,cm^{-2}\,s^{-1} = 1.71 \times 10^{-4}\,ph\,cm^{-2}\,s^{-1}$ \\
      40-100 keV: $1\,mCrab = 9.42 \times 10^{-12}\,erg\,cm^{-2}\,s^{-1} = 9.67 \times 10^{-5}\,ph\,cm^{-2}\,s^{-1}$}\\
\tablefoottext{d}{Type identifiers: LMXB=Low Mass X-ray binary, HMXB=High Mass X-ray binary.}
}
\label{tab:microquasars}
\end{table*}

\begin{table*}
\caption{\label{t7}511 keV $2\,\sigma$ flux upper limits for the hard X (E $>$ 20~keV) Supernova Remnants (SNR) reported in the 3rd IBIS catalogue \citep{Bird2007}.}
\centering
\begin{tabular}{l r r c r r c}
\hline\hline
Source name & R.A. & Dec & Error\tablefootmark{a} &  $F_{20-40}$\tablefootmark{b} & $F_{40-100}$\tablefootmark{b} & $2\,\sigma$ flux limit \\
                      & (deg)  & (deg)  &  (arcminutes)  &  (mCrab)         &  mCrab            & ($10^{-4}\,ph\,cm^{-2}\,s^{-1}$) \\
\hline
4U 0022+63           &         6.319    &   64.159   &     3.6   &   $0.7 \pm 0.1$    &     $0.7 \pm 0.2$     &      2.3 \\
PSR J1811-1926    &    272.827     &  19.417    &    2.9    &  $0.8 \pm 0.1$     &    $1.0 \pm 0.2$      &    14  \\
IGR J18135-1751   &   273.395      & -17.871   &     2.2    &  $1.1 \pm 0.1$     &   $1.7 \pm 0.2$      &    2.3 \\
SNR 021.5-00.9      &   278.388      & -10.579   &     1.2    &  $3.3 \pm 0.1$     &    $3.3 \pm 0.2$     &     3.0 \\
AX J1838.0-0655  &   279.509     &   -6.916     &    1.5     &   $1.9 \pm 0.1$     &    $3.0 \pm 0.2$     &     2.9 \\
Kes 73                    &  280.338     & -4.948        &    1.3     &   $2.2 \pm 0.1$     &    $4.2 \pm 0.2$     &     2.9 \\
AX J1846.4-0258   &   281.596   &  -2.983       &      1.8    &  $1.7 \pm 0.1$     &    $2.4 \pm 0.2$      &    2.8 \\
Cas A                    &   350.848    & 58.815      &      0.9    &  $4.1 \pm 0.1$      &   $2.4 \pm 0.2$       &   2.5  \\
\hline
\end{tabular}
\tablefoot{\\
\tablefoottext{a}{Position error expressed as radius of 90 \% confidence circle in arcminutes.} \\
\tablefoottext{b}{Fluxes averaged on the total exposure: \\
      20-40 keV: $1\,mCrab = 7.57 \times 10^{-12}\,erg\,cm^{-2}\,s^{-1} = 1.71 \times 10^{-4}\,ph\,cm^{-2}\,s^{-1}$ \\
      40-100 keV: $1\,mCrab = 9.42 \times 10^{-12}\,erg\,cm^{-2}\,s^{-1} = 9.67 \times 10^{-5}\,ph\,cm^{-2}\,s^{-1}$}\\
}
\label{tab:snr}
\end{table*}

\subsection{ISGRI  background and effective area at 511 keV}
As discussed in the previous section, the IBIS {\em sensitivity} depends on the instrumental background and the detector's effective area at 511~keV. Since the background is caused by radiation induced by cosmic rays interactions with the satellite, the telescope, and the upper atmosphere, the background is anti-correlated with the solar activity, reaching a minimum at the solar max in the early periods of the INTEGRAL activity. During this period, we measured a 511~keV background\footnote{We define the 511 keV background as the count rate in a $FWHM$  width bin centered at 511~keV} of about $8\, counts/s$. The long-term background trend is characterized by a general increase with time and  a mean value of about $10\, counts/s$ (Figure \ref{fig:bkg}). The peaks in this curve are due to the high-energy component of solar flares, which produce bursts of radiation across the electromagnetic spectrum from radio waves to X and $\gamma$-rays. In our data sample,  we removed any time intervals associated to these bursting events, because they are affected by systematic noise,  so are not useful for scientific data analysis.

The ISGRI effective area is estimated by the Monte Carlo method (\S \ref{sec:mc}). A  mono-energetic 511~keV point source is obtained by simulating a parallel beam incident on the top of the detector. If $N_{in}$ is the number of input photons, the effective area is simply estimated by
\begin{equation}
A_{eff}(511\, keV) = A_{beam} \times \frac{N_{det}}{N_{in}} = (54.2 \pm 0.2) \, cm^{2} 
\label{area-isgri}
\end{equation}
where $A_{beam}$ is the area covered by the beam, and $N_{det}$ is the number of the photons detected in a $FWHM$  width bin centered at 511~keV. Taking  the ISGRI geometrical area ($\sim 3000\, cm^{2}$) into account, the effective area results from of  an about 2~\% peak efficiency at 511~keV.

\subsection{IBIS 511 keV line sensitivity}
   If the 511~keV flux measured by SPI was due to a set of point sources, would it be possible to detect some of these sources with IBIS? To answer this question we consider the simple case where all the flux is concentrated in just one point source.  In this case, we would detect a flux of $10^{-3}\, ph\, cm^{-2}\, s^{-1}$ with a count rate of
\begin{equation}
S_{511\, keV} = 10^{-3} \times \frac{A_{eff}}{2} \times eff_{img}=  (2.03 \pm 0.01) \times 10^{-2}\, s^{-1}
\end{equation} 
From equation (\ref{eq:sens}) it follows that for an exposure of $T = 10^{7}\, s$, assuming a background $B_{511\, keV} = 10\, counts/s $,  we obtain a signal to noise of
\begin{equation}
\frac{s}{n} = S_{511\, keV}  \times \sqrt{\frac{T}{B_{511\, keV}}} = 20\, \sigma
\label{eq:significance}
\end{equation} 
We conclude that for a 10 Ms exposure the $2\sigma$ 511~keV {\em sensitivity} is on the order of $10^{-4}\,ph\,cm^{-2}\,s^{-1}$.


   The IBIS sensitivity can also be estimated by using data from the Crab observations. The Crab Nebula was monitored by INTEGRAL through a large number of pointings that were performed during many dedicated observations.  The Crab exposure that we obtained with these data is equal to $3\, Ms$. Reducing all the available data, we detect in the sky mosaic the Crab Nebula from 20~keV up to 531~keV (Fig. \ref{fig:crab-gamma}). The signal goes from $8372\, \sigma$ in the $20-40$ keV band to $3.8\, \sigma$ in the $491-531$ keV (Table \ref{tab:crab}). As the last energy band is where we are looking for a direct positron annihilation signal from point sources in the data, this signal gives us the opportunity to estimate the IBIS sensitivity at 511~keV using real data. A recent Crab measurement made with the INTEGRAL/SPI $\gamma$-ray spectrometer is reported by  \cite{Jourdain2008}, where the data are modeled by a broken power law (Table \ref{tab:broken}). The $491$-$531$ keV Crab flux  is equal to  
\begin{equation}
F_{Crab}(491-531\, keV) = (6.8 \pm 0.5) \times 10^{-4}\, ph\, cm^{-2}\, s^{-1} 
\label{eq:fcrab511}
\end{equation}
Since we detect this flux at a significance level of $3.8\, \sigma$, the sensitivity at $2\, \sigma$ level is $3.6 \times 10^{-4}\, ph\, cm^{-2}\, s^{-1}$, obtained with an exposure of $3\, Ms$. With a $10\, Ms$ exposure the sensitivity (Equation \ref{eq:sens}) will be
\begin{equation}
S_{2\sigma}^{Crab} (511\, keV) =  (2.0 \pm 0.7) \times 10^{-4}\, ph\, cm^{-2}\, s^{-1} \\
\end{equation}

  
\subsection{The Galactic center at 511 keV}     
   With the only exception given by the Crab, detected at about 4 sigma of significance (Table \ref{tab:crab}), with about five years of data, we obtain a  511~keV (491-531~keV bin) all sky map of our Galaxy without finding any signal. We note that, as the Crab signal is consistent with the expected continuum contribution at this energy, we do not find any evidence of a 511~keV emission line in this source. At each point on this map the variance is given by the {\em Standard Analysis} of the raw data. Then, at a given position, dividing the standard deviation  into the detector effective area and imaging efficiency, we estimate the sensitivity at 511 keV.  In the Galactic center (Sgr A*) position, we estimate for an exposure of 10~Ms, the following flux upper limit (given at at $2\, \sigma$ of significance level)
\begin{equation}
S_{2\sigma}(Sgr A^*) = 1.6 \times 10^{-4}\,ph\,cm^{-2}\,s^{-1}  
\label{eq:sgralimit}
\end{equation}


   We obtain a similar 511~keV flux upper limit for the sources located in the Galactic bulge, where the observations have an exposure of about 10 Ms. We then focused our interest on the this region on different timescales. Due to the requirement on the orientation of the INTEGRAL solar panels, the region of the Galactic center can be observed in only two periods per year that we call spring and autumn as they occur mainly in these seasons in the northern hemisphere. A Ms time scale is explored by considering the observation of the Galactic center during these visibility periods. To select the data, we extracted from the data archive all the IBIS pointings with Sgr~A* in the field of view. We then obtain a 511~keV sky mosaic for each Galactic center visibility period. No significant excess in the 511~keV bin are detected. We report the 511~keV flux limits in the center of the Galaxy during the INTEGRAL visibility periods in Table \ref{tab:gcvis} . 
      
\subsection{511 keV flux upper limit for compact sources}
   The LMXBs detected by IBIS in the Galactic plane could make a substantial contribution to the galactic 511~keV emission \citep[see][]{Weidenspointner2008,Weidenspointner2008a}. Among a total of 79 LMXBs reported in the hard X domain by \cite{Bird2007}, 70 are located  near to the Galactic plane ($|b| < 10$), and then can in principle contribute to the Galactic 511~keV emission. For each of these sources, we evaluate the 511~keV flux limits. Depending on the exposure, the flux limit goes from $1.6 \times 10^{-4}\,ph\,cm^{-2}\,s^{-1}$ in the maximum exposure (10~Msec) sky regions to about $4.0 \times 10^{-4}\,ph\,cm^{-2}\,s^{-1}$ (Fig. \ref{fig:lmxb-hist}). The only exception is 4U~0614+091, for which we obtain a higher flux limit ($8.7 \times 10^{-4}\,ph\,cm^{-2}\,s^{-1}$), because it is located in a sky region with a lower exposure.
   
   Among X-ray binaries (XRBs), microquasars (MQs) are promising galactic positron emitters \cite[see][]{Guessoum2006}. We therefore report the 511~keV upper limits for the X-ray Binaries classified as microquasars and detected by IBIS above 20~keV (Table \ref{tab:microquasars}). Among the 10 MQs in our sample, 7 are  LMXBs, 3 High Mass X ray Binaries (HMXBs).

   Some of the atomic nuclei produced in supernovae produce positrons by  radioactive $\beta^{+}$ decay of the nucleosynthesis products. The initial positron energy is expected $\sim\, 1\,MeV$, in agreement with the injection energy constraint \citep{Beacom2006}. The 511~keV flux upper limits obtained with our data analysis for the supernova remnants (SNRs) detected by IBIS above 20~keV are reported in Table \ref{tab:snr}.
These limits have been estimated by considering these objects as possible 511~keV point-like sources, i.e. with an extension lower than the IBIS angular resolution (12 arcminutes). However, the $\gamma$-ray emission from SNRs could be more extended, in which case the sensitivity degrades almost linearly with the source extent \citep{Renaud2006}.
   
    

\section{Discussion}
\label{discussion}
   We have focused data analysis on the search for 511 keV line emission from point sources. A possible detection at this energy would be a proof of an $e^{+}e^{-}$ annihilation process in Galactic sources, providing an important clue to the mystery on the origin of Galactic bulge positrons. Through a data analysis of about five years of observations, we do not detect any 511~keV $\gamma$-ray point source with IBIS. The flux upper limits that we report for the sources located in the Galactic bulge are on the order of $2 \times 10^{-4}\,ph\,cm^{-2}\,s^{-1}$. A direct consequence of this result is that, if the flux of positrons measured by SPI from the Galactic bulge ($\sim 10^{-3}\,ph\,cm^{-2}\,s^{-1}$) was caused just by point sources, assuming the same intensity for all of them,  there should be at least six sources. As the sensitivity degrades with energy, a better constraint on the 511~keV emission should be achieved by focusing data analysis to the lower energy continuum emission due to ortho-positronium. However, although the value of fraction of positronium measured by SPI for the diffuse Galactic bulge emission is close to one, the annihilation radiation from Galactic objects may bias the measurements. In this regard, we note that the SPI data analysis is optimized for narrow gamma lines with poor sensitivity to broad emission lines. In terms of data analysis, the positronium continuum needs to be separated by the intrinsic source continuum emission up to a few hundred keV. Therefore, to avoid contamination, a simple imaging technique can use an energy bin with just a fraction of the positronium continuum due to positronium, thereby degrading the sensitivity that should be obtained with this method. A different approach, based on spectral analysis, may be to search for any positronium components in the continuum emission of the Galactic sources.


   
Taking the Galactic center distance into account, the measured 511~keV flux corresponds to the annihilation of $2 \times10^{43}$ positrons per second. We note that, since the total X-ray luminosity estimated in LMXBs is $\sim 10^{39}\, erg/s$ \citep{Grimm2002}, the energy ($2 \times 10^{37}\, erg/s$) required to create the Galactic positrons can in principle come from these sources. The idea that these sources could make an important contribution to the annihilation $e^{+}e^{-}$ radiation is also supported by the correlation between their spatial distribution and the distribution of the 511~keV radiation measured by SPI \citep{Weidenspointner2008}. The fact that we do not detect any LMXB with IBIS is consistent with the idea that a significant part of the positrons propagates in the interstellar medium before being annihilated away from the birth place. For the microquasars, a possible 511~keV annihilation radiation can arise near the source, from the interaction of the jet with the interstellar medium, or  with the companion star in the case of a misaligned jet \citep{Guessoum2006}. In general, a possible electron-positron annihilation emission due to a Galactic jet would be seen as a point source if its angular size is lower than 12 arcminutes (i.e. the IBIS angular resolution). The study of the jets from microquasars makes use of radio and X-ray observations \citep[for a recent review see][]{Gallo2010}. The jet interpretation of the radio emission was confirmed by the milliarcseconds scale jets observed in Cygnus X-1 \citep{Stirling2001} and GRS 1915+105 \citep{Dhawan2000}. In X-ray band, arcsecond scale jets have been observed for the first time in SS 433 with the Chandra X-ray observatory \citep{Marshall2002}. For microquasars 1E1740.7-2942, a broad emission line detected by SIGMA has been reported \citep{Sunyaev1991}. However, since it was not confirmed by OSSE data \citep{Jung1995}, this measure is quite controversial. With a 5-year sky monitoring at 511~keV with IBIS, we do not detect this phenomenon, showing that, if true, it must have a rather low cycle of activity.


The 511~keV Galactic luminosity could be explained by positrons generated by $\beta^{+}$ decay of the nucleosynthesis products in supernovae. If a significant fraction of positrons are annihilated near the birth place, \cite{Martin2010} have proposed that supernovae could be detected as 511~keV point sources, but the flux is expected to be lower than the actual instrument {\em sensitivity} at this energy. Depending on the temperature of the environment where the annihilation takes place, a 511~keV broad emission line cannot be ruled out for supernovae. If the process takes place in the hot medium like the shocked regions of Supernova remnants, the expected line broadening would be $\sim 30$~keV for a temperature of $10^{7}\, K$, and then the emission would be within the energy bin (491-531~keV) used in our data analysis.

\begin{acknowledgements}
I am grateful to Steven Shore and Mark Leising for useful comments on this paper. I am also grateful to Peter von Ballmoos, Julien Malzac, Brian McBreen, Giorgio Palumbo, Luigi Stella, and Pietro Ubertini for useful discussions. I acknowledge financial contribution from the agreement ASI-INAF I/009/10/0.
\end{acknowledgements}

\bibliographystyle{aa}
\bibliography{biblio}

\end{document}